# Financial Market Dynamics: Superdiffusive or not?

Sandhya Devi[†]

Edmonds, WA, 98020, USA

Email: sdevi@entropicdynamics.com

**Abstract:** The behavior of stock market returns over a period of 1-60 days has been investigated for S&P 500 and Nasdaq within the framework of nonextensive Tsallis statistics. Even for such long terms, the distributions of the returns are non-Gaussian. They have fat tails indicating that the stock returns do not follow a random walk model. In this work, a good fit to a Tsallis q-Gaussian distribution is obtained for the distributions of all the returns using the method of Maximum Likelihood Estimate. For all the regions of data considered, the values of the scaling parameter $q$, estimated from one day returns, lie in the range 1.4 to 1.65. The estimated inverse mean square deviations (*beta*) show a power law behavior in time with exponent values between $-0.91$ and $-1.1$ indicating normal to mildly subdiffusive behavior. Quite often, the dynamics of market return distributions is modelled by a Fokker-Plank (FP) equation either with a linear drift and a nonlinear diffusion term or with just a nonlinear diffusion term. Both of these cases support a q-Gaussian distribution as a solution. The distributions obtained from current estimated parameters are compared with the solutions of the FP equations. For negligible drift term, the inverse mean square deviations (*beta*$_{FP}$) from the FP model follow a power law with exponent values between $-1.25$ and $-1.48$ indicating superdiffusion. When the drift term is non-negligible, the corresponding *beta*$_{FP}$ do not follow a power law and become stationary after certain characteristic times that depend on the values of the drift parameter and $q$. Neither of these behaviors is supported by the results of the empirical fit.

## 1. Introduction

Many well-known financial models [1] are based on the efficient market hypothesis [2] according to which: a) investors have all the information available to them and they independently make rational decisions using this information, b) the market reacts to all the information available reaching equilibrium quickly, and c) in this equilibrium state the market essentially follows a random walk [3]. In such a system, extreme changes are very rare. In reality however, the market is a complex system that is the result of decisions by interacting agents (e.g., herding behavior), traders who speculate and/or act impulsively on little news, etc. Such a collective/chaotic behavior can lead to wild swings in the system, driving it away from equilibrium into the realm of nonlinearity, resulting in a variety of interesting phenomena such as phase transition, critical phenomena such as bubbles, crashes [4], superdiffusion [5] and so on.

---

[†] Shell International Exploration and Production Co. (Retired)

The entropy of an equilibrium system following a random walk is given by Shannon entropy [6]. Maximization of this entropy [7] with constraints on the first three moments yields a Gaussian distribution. Therefore, if the stock market follows a pattern of random walk, the corresponding returns should show a Gaussian distribution. However, it is well known [8] that stock market returns, in general, show a more complicated distribution. This is illustrated in Figure 1 which compares the distributions of 1 day and 20 day log returns of S&P 500 and Nasdaq stock markets (1994-2014) with the corresponding Gaussian distributions. The data distributions show sharp peaks in the center and fat tails over many scales, neither of which is captured by the Gaussian distribution. Several studies [9] [10] indicate that these issues can be addressed using statistical methods based on Tsallis entropy [11], which is a generalization of Shannon entropy to nonextensive systems. These methods were originally proposed to study classical and quantum chaos, physical systems far from equilibrium such as turbulent systems, and long range interacting Hamiltonian systems. However, in the last several years, there has been considerable interest in applying these methods to analyze financial market dynamics as well. Such applications fall into the category of econophysics [5].

In the random walk model, the dynamics of stock market returns are assumed to be described by the standard Ito-Langevin equation which has a linear (in time) drift term and a $\delta$ function correlated noise term that follows a Wiener process. The dynamics of the corresponding probabilities are described by a Fokker Plank (FP) equation with linear drift and diffusion terms. (It can be proved using standard techniques that the Ito-Langevin and Fokker-Plank equations are equivalent [12]). The solution to this FP equation is a Gaussian distribution. In this case, the autocorrelation of the standard deviation $\sigma$ falls as $\sqrt{\tau}$ where $\tau$ is the time delay. However, empirical evidence shows [5] that the autocorrelation function of $\sigma$, even for low frequency data, falls as $\tau^\gamma$ where $\gamma$ is $< 0.5$ indicating long range correlation. A generalization of Fokker-Plank equation that takes into account long-term correlation by using non-extensive statistical methods based on Tsallis entropy has been given by L. Borland [13]. This involves replacing the noise term in Ito-Langevin equation by a non-linear noise term that depends on some power of the probability. This results in a Fokker-Plank equation with a non-linear diffusion term. The solution to this equation, under some assumptions, is a Tsallis q-Gaussian distribution [14] [15]. The generalized inverse mean square deviation of this distribution follows a power law, with the magnitude of the exponent $> 1$. This points to a super diffusive process. This model has been applied to study both high frequency [16] [17] and long-term low frequency returns [18].

In reality, are the stock market returns superdiffusive? Analysis of very short-term stock returns (1-60 minutes) shows [16] that these high frequency returns are indeed superdiffusive. However, there have been several works [18] extending the superdiffusive model to longer-term stock returns (1 day – several months) and in particular option pricing. Figure 1 shows that even longer-term returns have fat tails. However, this by itself does not necessarily imply superdiffusion. The objective of this paper is to investigate empirically the assumptions made in the model discussed above and to test whether the distributions of low frequency stock returns show a super diffusive character. To do this, the parameters of the q-Gaussian distribution are estimated from the distributions of observed stock returns at different time delays. These, in particular, the generalized inverse mean square deviations ($\beta$), are then compared, at the same delays, with the $\beta$ given by the analytical solutions of the non-linear Fokker-Planck equations described above.

Computing the Tsallis distribution of returns involves accurate estimation of its parameters. The usual way is to fit the Tsallis distribution to the binned data distribution using a combination of linear regression and least square optimization techniques [19] [20] [21]. The tail regions of the data (Figure 1), which are important in the parameter estimation, have relatively fewer samples and this is further reduced by binning. Statisticians [22] have long applied Maximum Likelihood Estimation (MLE) method to estimate the parameters of the Pareto distribution, which for certain parameter choices gives a q-distribution. A comparison of several optimization techniques carried out by Clauset et al. [23] on synthetic data (following a power law distribution) shows that, for discrete data, MLE estimates give values closest to the real values. As shown and discussed in several references [24] [25], under some general conditions, MLE is a consistent estimator, in the sense that for large number of samples N, the estimated parameters approach the true values in a probabilistic sense. It is asymptotically normal, unbiased and consistent, which means that the distribution of errors between the estimated and true values is Gaussian with zero mean and covariance given by $I^{-1}/N$, where $I$ is the Fisher Information matrix. Further, the variance of the estimator attains the lower limit of Cramer-Rao inequality [26]. Hence, one can calculate the standard errors of the estimated parameters {i} as $\sqrt{I^{-1}_{ii}/N}$. Similar conclusions were drawn by Shalizi [27] who applied the method to q-exponential distributions. In this paper MLE is used to estimate the parameters of a q-Gaussian distribution.

The paper is organized as follows. In Section 2, a brief review of Tsallis entropy, the q-Gaussian distribution, and the non-linear Fokker-Planck equation for the evolution of probability density function will be given. The MLE equations for a q-Gaussian distribution will be discussed in Section 3. Section 4 deals with the application to market data (S&P 500 and Nasdaq) and comparisons of the estimated parameters and distributions at different time scales with those given by the solutions of the FP equation. Conclusions will be given in Section 5.

## 2. A Brief Review of the Theory

The Tsallis generalization of Shannon Entropy:

$$S_{sh} = \sum_i P_i \, ln(1/P_i) \qquad\qquad\qquad (1)$$

to nonextensive systems is given by:

$$S_q = \sum_i P_i \, ln_q(1/P_i) \tag{2}$$

where $P_i$ is the probability density function at the i$^{th}$ sample under the condition $\sum_i P_i = 1$ and the $q$ logarithm $ln_q(x)$ is given by

$$ln_q(x) = (x^{1-q} - 1)/(1-q) \tag{3}$$

$q$ is a universal parameter, but its value can change from system to system.

Substituting (3) in (2), we get:

$$S_q = \left(1 - \sum_i P_i^q\right)/(q-1) \tag{4}$$

It is important to note that unlike Shannon Entropy, Tsallis entropy is <u>not additive</u> which points to its applicability to correlated systems.

Considering the continuous case for a random variable $\Omega$, one can show [11] that the maximization of $S_q$ with respect to $P$ under the following constraints:

$$\int_{-\infty}^{\infty} P(\Omega)d\Omega = 1 \tag{5a}$$

$$\langle(\Omega - \bar{\Omega})\rangle_q = \int_{-\infty}^{\infty}(\Omega - \bar{\Omega}) \; P^q(\Omega)d\Omega = 0 \tag{5b}$$

$$\langle(\Omega - \bar{\Omega})^2\rangle_q = \int_{-\infty}^{\infty}(\Omega - \bar{\Omega})^2 \; P^q(\Omega)d\Omega = \sigma_q^2 \tag{5c}$$

gives the Tsallis distribution:

$$P_q(\Omega) = \frac{1}{Z_q} \, [1 + (q-1)\beta(\Omega - \bar{\Omega})^2]^{1/(1-q)} \tag{6}$$

$Z_q$ is the normalization given by:

$$Z_q = \int[1 + (q-1)\beta(\Omega - \bar{\Omega})^2]^{1/(1-q)} \, d\Omega \tag{7}$$

Here $\beta$ is the Lagrange multiplier of the constraint (5c) and is given by:

$$\beta = 1/(2\sigma_q^2 \, Z_q^{q-1})$$

It is straightforward to show that:

$$Z_q = C_q/\sqrt{\beta} \tag{8}$$

$$C_q = \sqrt{\pi} \frac{\Gamma\left(\frac{1}{q-1} - \frac{1}{2}\right)}{\sqrt{q-1}\ \Gamma\left(\frac{1}{q-1}\right)} \tag{9}$$

Here $\overline{\Omega}$ is the mean value of $\{\Omega_i\}$. $\Gamma$ is the gamma function. Note that:

a) In the limit $q \to 1$, it can be shown that the Tsallis entropy and the corresponding distribution go to the Shannon entropy and the Gaussian distribution respectively.

b) Unlike the Gaussian distribution case, the regular variance is not defined for all $q$. It is given by:

$$\sigma^2 = 1/(5 - 3q)\beta \qquad\qquad q < 5/3 \tag{10}$$

Let us now look at the evolution of $P_q(\Omega)$ across time scales. It has been shown [14] [15] that a solution to a nonlinear diffusion (Fokker-Plank) equation of the form:

$$\partial P(\Omega,t)/\partial t = -\partial\left[f(\Omega)P(\Omega,t))\right]/\partial\Omega + \frac{D}{2}\ \partial^2 P(\Omega,t)^\nu/\partial\Omega^2 \tag{11}$$

is:

$$P(\Omega,t) = \frac{1}{Z_q(t)}\left[1 + (q-1)\beta(t)(\Omega - \overline{\Omega})^2\right]^{1/(1-q)} \tag{12}$$

Here the drift function term $f(\Omega)$ is assumed to be:

$$f(\Omega) = a - b\Omega$$

The probability density function (PDF) given by (12) satisfies (11) under the following conditions:

$$q = 2 - \nu$$

$$\left(\frac{1}{\nu+1}\right)\partial Z_q^{\nu+1}/\partial t + bZ_q^{\nu+1} - 2D\nu\left(\beta(0)Z_q^2(0)\right) = 0 \tag{13}$$

$$\left[Z_q(t)/Z_q(0)\right]^2 = \beta(0)/\beta(t) \tag{14}$$

$$\frac{d\overline{\Omega}}{dt} = a - b\,\overline{\Omega} \tag{15}$$

From (12) – (15), it is straightforward to show:

$$\beta(t)^{-(3-q)/2} = [2\,(2-q)D/b]\,C_q^{(q-1)/2}\big[1 - e^{-t/\tau}\big] \tag{16}$$

Here $\tau = 1/(b(3-q))$ is the characteristic time and $C_q = \beta(0)\,Z_q^2(0)$ is constant in time. A comparison of (14) with (8) shows that the norm is conserved. In deriving (12) – (16), a boundary condition $P(\Omega, 0) = \delta(\Omega)$ (implies $\beta(0) = \infty$) is used.

If the drift term is negligible ($b \to 0$), $t \ll \tau$, the exponential in (16) can be expanded up to linear term. In this case, $\beta$ is given by:

$$\beta(t) \propto t^{-2/(3-q)} \tag{17}$$

independent of the drift parameter $b$.

For $q > 1$, the absolute value of the exponent of $t$ in (17) is greater than 1. This means that the mean square deviation $(1/\beta)$ of $\Omega$ follows a power law in time with exponent greater than 1. In an anomalously diffusive system, the mean square deviation scales as $t^\eta$. It is superdiffusive if $\eta > 1$, subdiffusive if $\eta < 1$, and normal if $\eta = 1$. Therefore, according to the analysis above, for negligible drift term, the stock market returns should show a superdiffusive character. We will denote the $\beta$ for superdiffusion and drift + diffusion cases as $\beta_{sd}$ and $\beta_{dd}$ respectively.

To summarize, according to the model discussed above, the distributions of stock market returns are q-Gaussians at all delays and the market is superdiffusive, provided $q$ does not vary with time. Whether the market is superdiffusive under these assumptions will be investigated in the next sections.

## 3. Maximum Likelihood Estimation for q-Gaussian Distribution

### 3.1 Parameter estimation

In the Maximum Likelihood Estimation method, the parameters of a probability density function $P$, having N samples, are estimated by maximizing the objective function:

$$F = \sum_i ln(P_i) \tag{18}$$

For q-Gaussian distribution:

$$F = -N \, ln(Z_q) + (1/(1-q)) \, ln\left[\sum_i (1 + (q-1)\beta \, \Omega_i^2)\right] \tag{19}$$

Here, the variable $\Omega$ is assumed to be standardized. Making a change of variables:

$$\alpha = 1/(q-1), \qquad \kappa = \beta/\alpha \tag{20}$$

the objective function becomes:

$$F = -N \, ln(Z_q) - \alpha \, ln\left[\sum_i (1 + \kappa \, \Omega_i^2)\right] \tag{21}$$

where the normalization $Z_q$ in terms of the new variables is given by:

$$Z_q = \sqrt{(\pi/\kappa)} \, \left(\Gamma(\alpha - 1/2)/\Gamma(\alpha)\right) \tag{22}$$

Maximizing $F$ with respect to $\alpha$ and $\kappa$ gives:

$$[\psi(\hat{\alpha}) - \psi(\hat{\alpha} - 1/2)] = \overline{\log(1 + \hat{\kappa} \, \Omega^2)} \tag{23a}$$

$$1/2\hat{\kappa} = \hat{\alpha} \left(\overline{\Omega^2/(1 + \hat{\kappa} \, \Omega^2)}\right) \tag{23b}$$

Here $\psi$ is the digamma function. The bar denotes the mean value. $\hat{\alpha}$ and $\hat{\kappa}$ denote the estimated values of $\alpha$ and $\kappa$. In the limit $q \to 1$, (23b) gives $1/\beta = 2 \, \overline{\Omega^2}$.

Since (23b) depends on $\alpha$ explicitly, it can be eliminated from (23a), so that:

$$[\psi(f(\hat{\kappa})) - \psi(f(\hat{\kappa}) - 1/2)] = \overline{\log(1 + \kappa \, \Omega^2)} \tag{24}$$

where:

$$f(\hat{\kappa}) = (1/2\hat{\kappa}) \left[\overline{\Omega_i^2/(1 + \hat{\kappa} \, \Omega_i^2)}\right]^{-1}$$

Note that (24) depends only on $\hat{\kappa}$. But it is nonlinear and hence has to be solved numerically. Once $\hat{\kappa}$ is estimated using (24), $\hat{\alpha}$ can be estimated using (23b). The parameters $q$ and $\beta$ can then be computed from (20). We will denote the $q$ and $\beta$ so estimated by $q$ and $\hat{\beta}$.

In solving (23b) and (24), the range of $q$ is fixed between $1.1 - 1.66$ by requiring that we look for solutions with $q > 1$ and distributions with finite variance as given in (10). A reasonable initial guess for $1/\beta$ is the variance of the returns. For delays longer than 1 day, the initial guess for $1/\beta$ can be scaled as some function of the delay.

## 3.2 Error estimation

The errors in $\hat{\alpha}$ and $\hat{\kappa}$ (hence $q$ and $\hat{\beta}$) estimates can be calculated using the Fisher Information matrix $I$ which can be either the measured information matrix:

$$I_{kl}^{(m)} = \sum_i \frac{\partial log(P_i)}{\partial \varphi_k} \frac{\partial log(P_i)}{\partial \varphi_l} \tag{25}$$

or the expectation value:

$$I_{kl}^{(e)} = \langle \frac{\partial log(P)}{\partial \varphi_k} \frac{\partial log(P)}{\partial \varphi_l} \rangle \tag{26}$$

Here, $\varphi_i$ ($i = 1 \dots \text{m}$) are the parameters of the distribution $P$ and the expectation value is taken with $P$. The standardized errors for parameter estimates are then given by the diagonal elements of $I^{-1}$ evaluated at the estimated values. Therefore, the errors $S$ in $\hat{\alpha}$ and $\hat{\kappa}$ are:

$$S(\hat{\alpha}) = \sqrt{I_{\hat{\alpha}\hat{\alpha}}^{-1}/N}$$

$$S(\hat{\kappa}) = \sqrt{I_{\hat{\kappa}\hat{\kappa}}^{-1}/N} \tag{27}$$

Note that $I^{(m)}$ is data dependent and $I^{(e)}$ is only model dependent. As shown in the Appendix:

$$I_{\alpha\kappa}^{(e)} = \begin{bmatrix} I_{\alpha\alpha} & I_{\alpha\kappa} \\ I_{\kappa\alpha} & I_{\kappa\kappa} \end{bmatrix}$$

where:

$$I_{\alpha\alpha} = \psi_1(\alpha - 1/2) - \psi_1(\alpha) \tag{28a}$$

$$I_{\alpha\kappa} = I_{\kappa\alpha} = \frac{1}{2\kappa\alpha} \tag{28b}$$

$$I_{\kappa\kappa} = \left(\frac{1}{4\kappa^2}\right)\frac{(2\alpha-1)}{(\alpha+1)} \tag{28c}$$

and $\psi_1$ is the tri-gamma function.

The errors in $q$ and $\hat{\beta}$ can be obtained from those of $\hat{\alpha}$ and $\hat{\kappa}$ using the transformations (20).

## 4. Results

The data chosen for our analysis are S&P 500 and Nasdaq daily (close of the day) stock prices. The stock prices, which are de-trended with CPI to remove inflation trends, are displayed in Figure 2. We will consider the period after 1991 (about a year before the time when electronic trading over the internet was launched), since the character of the stock price variation changes dramatically after that. The time series shows a non-stationary character with wild fluctuations. The data for analysis is divided into two regions bounded by vertical dotted lines. Regions 1 and 2 cover the dot-com bubble period and the crash of 2008 respectively. Region 3 is reserved for testing and prediction purposes. This paper deals with the analysis of regions 1 and 2 only

The variables used for the estimation of $q$ and $\beta$ are the standardized log returns $\Omega(t, t_0)$ for delay $t$:

$$\Omega(t, t_0) = \left(y(t, t_0) - \mu_t\right)/\sigma_1 \tag{29}$$

computed for several starting times $t_0$ over the period of interest. Here

$$y(t, t_0) = \log(S(t_0 + t)) - \log(S(t_0))$$

S is the stock value, $\mu_t$ is the mean of $y(t)$, and $\sigma_1$ is the standard deviation for 1 day log returns. With this choice, $\bar{\Omega} = 0$. As discussed in Section 2, $q$ and $\beta$ are both estimated from 1 day standardized log returns. For delays greater than 1, $q$ is kept constant and only $\beta$ is estimated so that a comparison can be made with the solutions of the FP equations (11). The errors in $q$ and $\hat{\beta}$ are calculated using (28) and the transformation (20). For comparison, the errors from the measured Fisher Information matrix were also computed. The difference in errors in the parameters from the two methods is less than 0.3%.

### 4.1 Goodness of fit

Figures 3 and 4 show the comparison of the Tsallis distributions, from the estimated parameters, with the data distributions for regions 1 and 2 respectively. Also shown are the corresponding Gaussian distributions. To see how good the estimates are, Kolmogorov-Smirnov (KS) tests [28] [23] are performed at all delays considered. To do this, synthetic data are generated at each delay using a generalized Box-Müller method for generating q-Gaussian random deviates [29] given $q, \beta$ values. The synthetic data are standardized in the same way as the empirical data. Two types of tests are conducted.

a) The maximum absolute distances $\mathbf{D_{max}}$ between the empirical and synthetic cumulative distribution functions (CDF) are calculated. If $\mathbf{D_{max}}$ exceeds a critical distance $\mathbf{D_{crit}}$ [30] at a particular significance level, that fit should either be rejected or accepted at a higher significance level. $\mathbf{D_{crit}}$ is given by

$$\mathbf{D_{crit}} = c(\gamma) \sqrt{(n1 + n2)/(n1 * n2)}$$

Here, *n1* and *n2* are the number of samples in the empirical and synthetic CDF's respectively. The table for function $c(\gamma)$ at different significance levels $\gamma$ can be found in [30].

Figure 5 shows $\mathbf{D_{max}}$ as a function of delay. Also shown are the critical distances $\mathbf{D_{crit}}$ for a significance level of 0.05 (confidence 95%). All distances except for a delay around 30 days for Nasdaq are below the corresponding $\mathbf{D_{crit}}$. This value has to be accepted at a higher significance level of 0.10.

b) In the second test (as described in [23]), the number of points in the empirical CDF that are closer to the model than the corresponding synthetic CDF are calculated. The ratio of this number to the total number of points in the CDF gives a $\boldsymbol{P}$ value. If this value falls below the critical value $\boldsymbol{P_{crit}} = 0.1$, the fit is not considered good. Figure 6 shows plots of $\boldsymbol{P}$ as a function of delay. Except for a few isolated delays in the case of S&P 500 in region 2, all the $\boldsymbol{P}$ values are higher than the critical value.

In general, the distances for S&P 500 region 1 are much lower and the $\boldsymbol{P}$ values higher than those for the other 3 data. Also, the distances and $\boldsymbol{P}$ get worse with delay. One possibility for this is asymmetry. Note that, as the delay increases, the distributions (Figures 3 and 4) get more skewed towards large negative returns. But our model distribution is symmetric. The data for large positive returns is much sparser than that for large negative returns. Hence a better fit is obtained for large negative returns. This, on the average, could also lead to higher distances and lower $\boldsymbol{P}$ values. The asymmetry issue will be discussed more in subsection 4.4.

## 4.2 Tail index

If the CDF of a random variable $x$ follows a power law asymptotically

$$\text{CDF} \propto x^{-\eta}$$

then the exponent $\eta$ is called the tail index. In the case of q-Gaussian distribution, the

$$\text{CDF}_q \propto x^{-(q+1)/(q-1)}$$

when $(q-1)\beta x \gg 1$. Hence for $q > 1$, $(q+1)/(q-1)$ is the tail index. In the literature, tail indices are reported for several stock market returns. A few will be mentioned here. Gopikrishnan et al. [31] report a value of ~3 for high frequency (of the order of minutes)

S&P 500 returns over the period 1994–1995. Jiang et al. [32] estimate η to be between 2.78 and 4 for very high frequency Chinese stock returns for 2003. An analysis of low frequency (quarterly) returns for US, some European and emerging markets by Jondeau and Rockinger [33] over the period of 1965–2002 yields tail index values between 3 and 5 for negative (-ve) returns and 3 and 7 for the positive (+ve) returns. In general, the values of η seem to lie between 2 and 7. Using the relationship mentioned above between the tail index and q, we get comparable tail index values of 6, 4.77, 4 and 4.92 respectively for the two markets and the two regions considered.

## 4.3 Variation of $\beta$ with delay

The estimated values of $q$ (given at the top of Figures 3 and 4) are greater than 1 in all cases pointing to the non-Gaussian character. The $q$ values are different for each region indicating the change in the character of the data. The S&P 500 region 1 has the lowest $q$ of all the four data sets. The value of $q$ significantly depends on the tail characteristics of the distributions. For region 1 (dot-com bubble), the wilder swings of Nasdaq returns (Figure 2) result in fatter tails yielding a higher value of $q$. Similarly, for region 2, both S&P 500 and Nasdaq have fatter tails during the crash period resulting in higher values of $q$. Such variations in $q$ from region to region is also observed in [21]. This brings us to question the assumption of constant $q$. For the present analysis, we have dealt with this issue by splitting the data into different time regions. However, more rigorous investigations are needed since one cannot predict when the market characteristics will change from one of relative calmness to one of wild changes.

The variation of $\hat{\beta}$ with the delay $t$, along with error bars, are shown in Figure 7 on a log-log scale. The error in $\hat{\beta}$ is largest (~5%) for $t = 1$, when both $q$ and $\beta$ are estimated. For other values of $t$ it is less than 3%. The straight line character of the plots shows that

$$\hat{\beta} \propto t^{\lambda}$$

with $\lambda$ between $-0.91$ and $-1.1$. This points to a normal to mildly subdiffusive behavior.

A comparison of $\hat{\beta}$ with $\beta_{dd}$ and $\beta_{sd}$ is shown in Figure 8. Note that the computation of $\beta_{dd}$ depends on the drift parameter $b$ and the diffusion parameter D. These were estimated as follows. The drift parameter $b$ was estimated by fitting the ratio $\hat{\beta}(t)/\hat{\beta}(1)$ to the corresponding ratio of $\beta_{dd}$. Once $b$ is estimated, the diffusion parameter D is obtained by setting $\beta_{dd}(1) = \hat{\beta}(1)$. The values of $b$ and D and the corresponding characteristic times $\tau = 1/(b(3-q))$ are given in Table 1. For values of $t < \tau$, $\beta_{dd}$ shows an almost power law behavior with an exponent value less than that of $\beta_{sd}$ and closer to that of $\hat{\beta}$. However, for $t > \tau$, $\beta_{dd}$ changes its slope and approaches a stationary value. Therefore $\tau$ should be considered as the upper time limit for the validity of the drift + diffusion model.

Comparisons of the distributions of data with Tsallis distributions computed with $\hat{\beta}$, $\beta_{dd}$ and $\beta_{sd}$ are shown in Figures 9 and 10. Note that the superdiffusion and drift + diffusion curves are calculated from equations (12) and (16) using the estimated $q$ values. For smaller delays, there is good agreement between all the model distributions and the data. However, as the delay increases, the distributions from both the drift + diffusion and the superdiffusion models start deviating from the empirical fit and the data distributions, with the superdiffusion model deviating the most both for small and large returns.

### 4.4 Asymmetry

As seen in Figures 3 and 4, the distributions get more asymmetric with delay, the left side (-ve returns) getting fatter tails than the right side (+ve returns). However, the q-Gaussian model which is fit to the data is symmetric. Does this asymmetry affect the conclusion that the data shows normal diffusive behavior? To test this, the parameters $q$ and $\beta$ were estimated separately for the left and right branches of the distributions. Note that the estimation of $q$ is carried out from the distributions of one day returns where the asymmetry is not significant. Hence the estimated q-values in the asymmetric case are only about 1 – 4% different from the q-values in the symmetric case. Figure 11 shows the variation of estimated $\hat{\beta}$ with the delays, on a log-log scale, for the two branches of the distribution. For S&P 500 region 1, $\hat{\beta}_+$ for the right branch (+ve returns) are very close to those for left branch ($\hat{\beta}_-$) indicating less asymmetry. In the other three cases, $\hat{\beta}_+$ is shifted higher, indicating the right branch of the distributions has smaller width. This is also borne out from Figures 3 and 4. However, the log-log plots of both $\hat{\beta}_+$ and $\hat{\beta}_-$ still show a straight line character, with slopes close to -1, pointing to normal diffusive behavior of market returns.

## 5. Conclusions

Investigations of the behavior of the S&P 500 and Nasdaq stock market long-term returns, over a period which includes both the dot-com bubble of 2000 and the crash of 2008, show that the distributions of the returns are non-Gaussian and fat-tailed even for as long a term as 1-60 days. The distributions can be modelled well with a Tsallis q-Gaussian distribution, the parameters $(q, \beta)$ of which have been estimated using the Maximum Likelihood Estimation method. The values of $q$ are greater than 1 for all the regions considered, with high values for the dot-com bubble and the crash of 2008 periods. However, the inverse mean square deviation $\beta$ shows a power law behavior with exponent value very close to $-1$.

In several earlier works generalizing market returns to non-Gaussian distributions [18], the dynamics is assumed to be described by a nonlinear Fokker-Plank equation with only a nonlinear diffusion term. A solution to this equation is a Tsallis distribution. In this model, the $\beta$ variation, for a constant $q > 1$, follows a power law in time with the magnitude of the exponent greater than 1, pointing to superdiffusion. However, as discussed above, the present analysis of long-term market returns shows that, even though the distributions can be modelled with a Tsallis distribution with $q > 1$, the parameter $\beta$ falls approximately as $1/t$, indicating normal diffusion.

In fact, as the time delay increases, the distributions computed from the superdiffusion model deviate considerably from the corresponding data distributions.

The FP equation (11) supports a Tsallis q-Gaussian distribution as a solution when a drift term is included in addition to the diffusion term. But the variation of $\beta$ with time is not a power law. In addition, it approaches a stationary value for times greater than the characteristic time $\tau = 1/(b * (3 - q))$. It should however be noted that for $t < \tau$, the model with the drift + diffusion terms yields distributions that agree with the data distributions better than those from superdiffusion model.

As the delay increases, the distributions become increasingly asymmetric. However, our preliminary tests, by estimating the parameters separately for the +ve and -ve returns branches of the distributions, show that asymmetry does not change the conclusion that the market returns are almost normal diffusive.

The variation of the fitted values of $q$ from region to region throws doubt on the assumption of constant $q$. In the present work, this has been dealt with in an ad hoc manner by breaking the data into different time regions. More rigorous investigations are needed in this respect.

The present investigations show that the stock market dynamics, for longer delays such as considered in the present work, cannot be adequately modelled with a Fokker-Plank equation that has a linear drift and a nonlinear diffusion term as given in (11). What is needed is a dynamical equation that yields solution close to Tsallis distribution, but shows normal diffusion.

## Acknowledgements

Many thanks to Sherman Page for a critical reading of the manuscript.

## Appendix: Expected Fisher Information Matrix for q-Gaussian PDF

In terms of the transformed parameters $\alpha$ and $\kappa$ given in (20), the expected Fisher Information matrix (26) is given by:

$$
\begin{aligned}
I_{\alpha\kappa}^{(e)} &= \begin{bmatrix} \langle \frac{\partial log(P)}{\partial \alpha} \frac{\partial log(P)}{\partial \alpha} \rangle & \langle \frac{\partial log(P)}{\partial \alpha} \frac{\partial log(P)}{\partial \kappa} \rangle \\ \langle \frac{\partial log(P)}{\partial \kappa} \frac{\partial log(P)}{\partial \alpha} \rangle & \langle \frac{\partial log(P)}{\partial \kappa} \frac{\partial log(P)}{\partial \kappa} \rangle \end{bmatrix} \\[2em]
&= -\begin{bmatrix} \langle \frac{\partial^2 log(P)}{\partial \alpha^2} \rangle & \langle \frac{\partial^2 log(P)}{\partial \alpha \partial \kappa} \rangle \\ \langle \frac{\partial^2 log(P)}{\partial \kappa \partial \alpha} \rangle & \langle \frac{\partial^2 log(P)}{\partial \kappa^2} \rangle \end{bmatrix}
\end{aligned}
\tag{A1}
$$

Using (6) – (9) and (20) and noting that $P$ is normalized, it is straightforward to show that:

$$
I_{\alpha\alpha} = -\langle \frac{\partial^2 \log(P)}{\partial \alpha^2} \rangle = \psi_1(\alpha - 1/2) - \psi_1(\alpha)
\tag{A2}
$$

$$
I_{\alpha\kappa} = I_{\kappa\alpha} = -\langle \frac{\partial^2 log(P)}{\partial \alpha \partial \kappa} \rangle = \frac{1}{2\kappa\alpha}
\tag{A3}
$$

$$
I_{\kappa\kappa} = -\langle \frac{\partial^2 log(P)}{\partial \kappa^2} \rangle = \left( \frac{1}{4\kappa^2} \right) \frac{(2\alpha - 1)}{(\alpha + 1)}
\tag{A4}
$$

Here, $\psi_1$ is the tri-gamma function. In deriving (A3) and (A4), the following expectation values are needed:

$$
\langle \frac{\Omega^2}{(1 + \kappa \, \Omega^2)} \rangle = \frac{1}{2\kappa\alpha}
$$

$$
\langle \frac{\Omega^4}{(1 + \kappa \, \Omega^2)^2} \rangle = \left( \frac{3}{4\kappa^2} \right) \frac{1}{\alpha(\alpha + 1)}
$$

The Fisher matrix $I_{q\beta}^{(e)}$, needed to compute the standard errors in $q$ and $\beta$, can be obtained from $I_{\alpha\kappa}^{(e)}$ using the transformation:

$$
I_{q\beta}^{(e)} = J \, I_{\alpha\kappa}^{(e)} \, \tilde{J}
\tag{A5}
$$

where $J$ is the Jacobian. From (20), it is straightforward to show:

$$J = \begin{bmatrix} -\alpha^2 & \kappa\alpha \\ 0 & 1/\alpha \end{bmatrix} \tag{A6}$$

# Figures

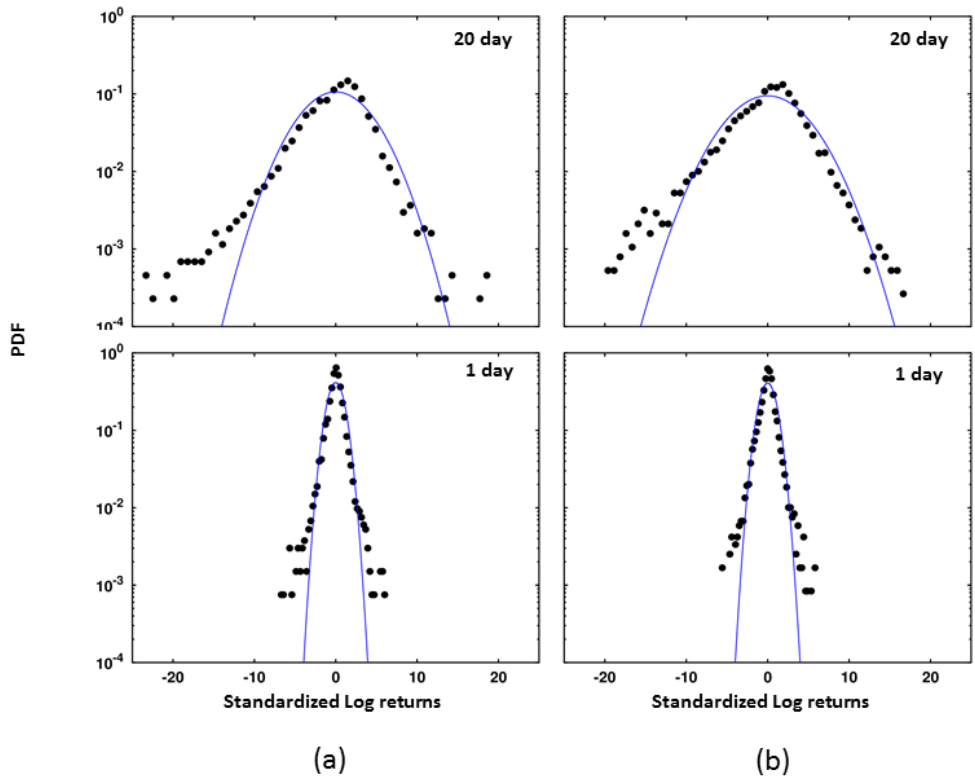

(a)                                      (b)

Figure 1.  Comparison of the distributions of standardized log returns (as given in (29)) with the Gaussian distributions (solid blue line) having the same mean and standard deviation as the data (black dots). (a) S&P 500 for 2 Jan 1994 – 31 Dec 2013 and (b) Nasdaq over the same period.

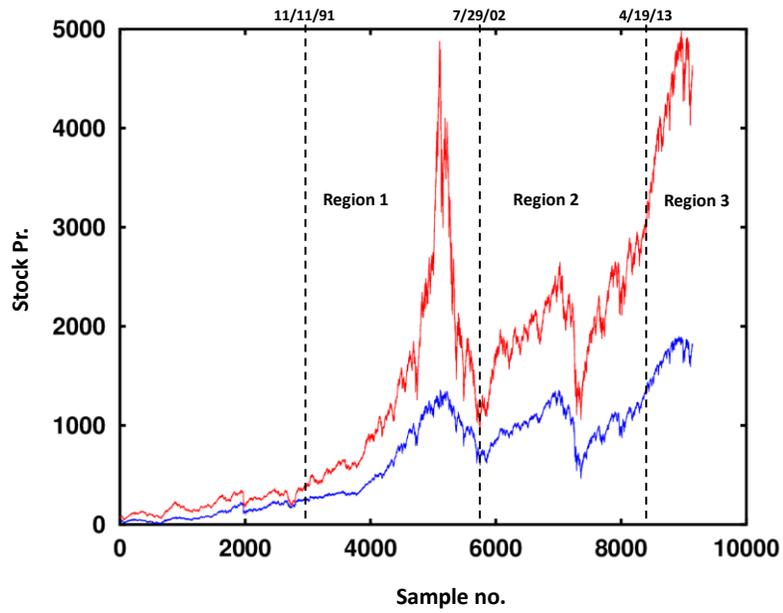

Figure 2.  S&P 500 and Nasdaq stock prices for 2 Jan 1980 – 31 March 2016. Region 1 (11 Nov 1991 – 29 Jul 2002) and region 2 (30 Jul 2002 – 4 Sep 2013) are chosen for analysis and region 3 (5 Sep 2013 – 31 Mar 2016) for testing.  Blue – S&P 500. Red – Nasdaq.

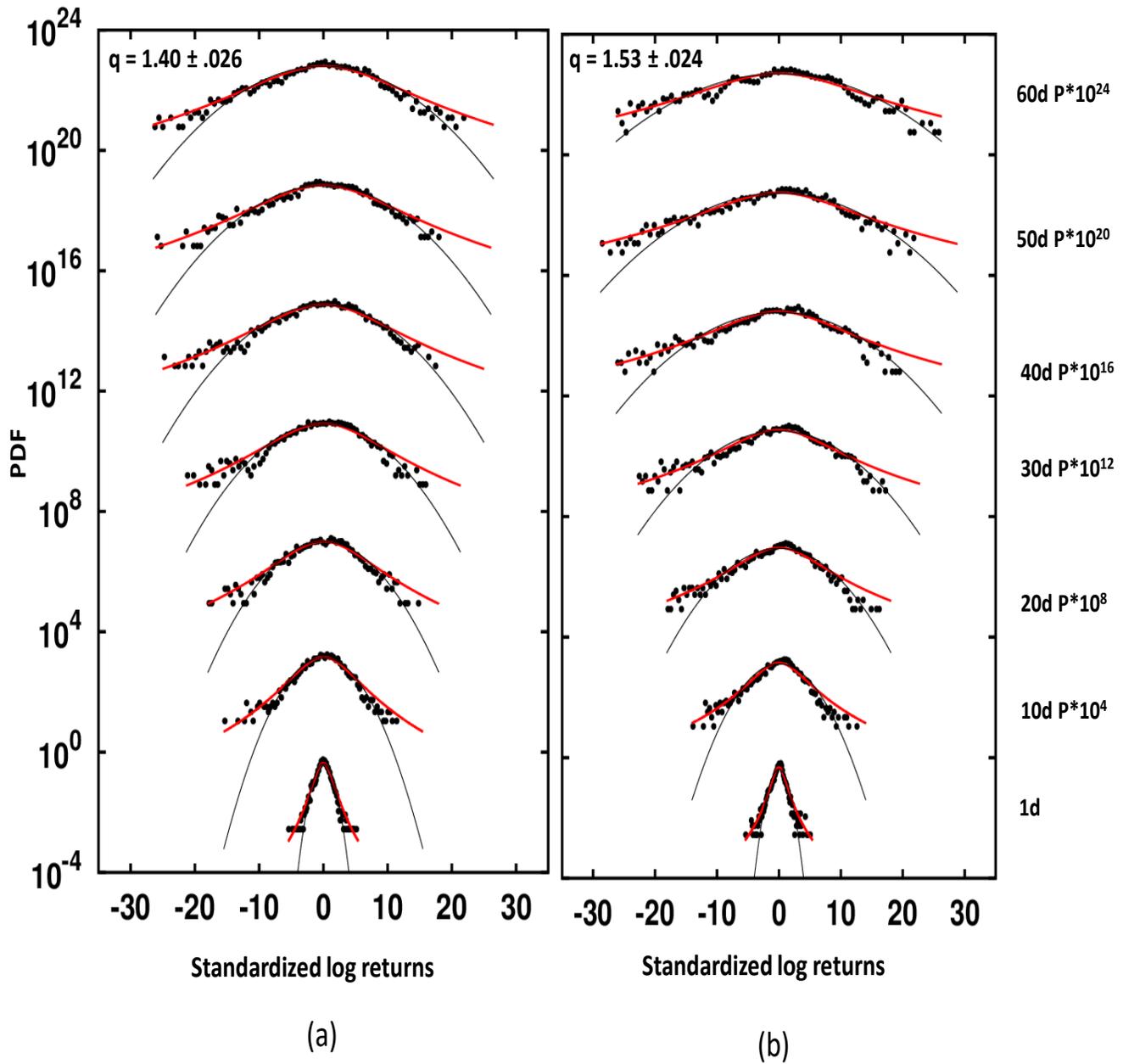

Figure 3. Comparison of the estimated Tsallis distributions with the data distributions for region 1 (11 Nov 1991 – 29 Jul 2002). Red – Estimated. Black – Gaussian. The delays corresponding to the distributions are given on the right hand side of the figure. The distributions for each delay are shifted by multiplying the corresponding PDF with the factors shown on the right hand side, next to the delays.  (a) S&P 500 and (b) Nasdaq.

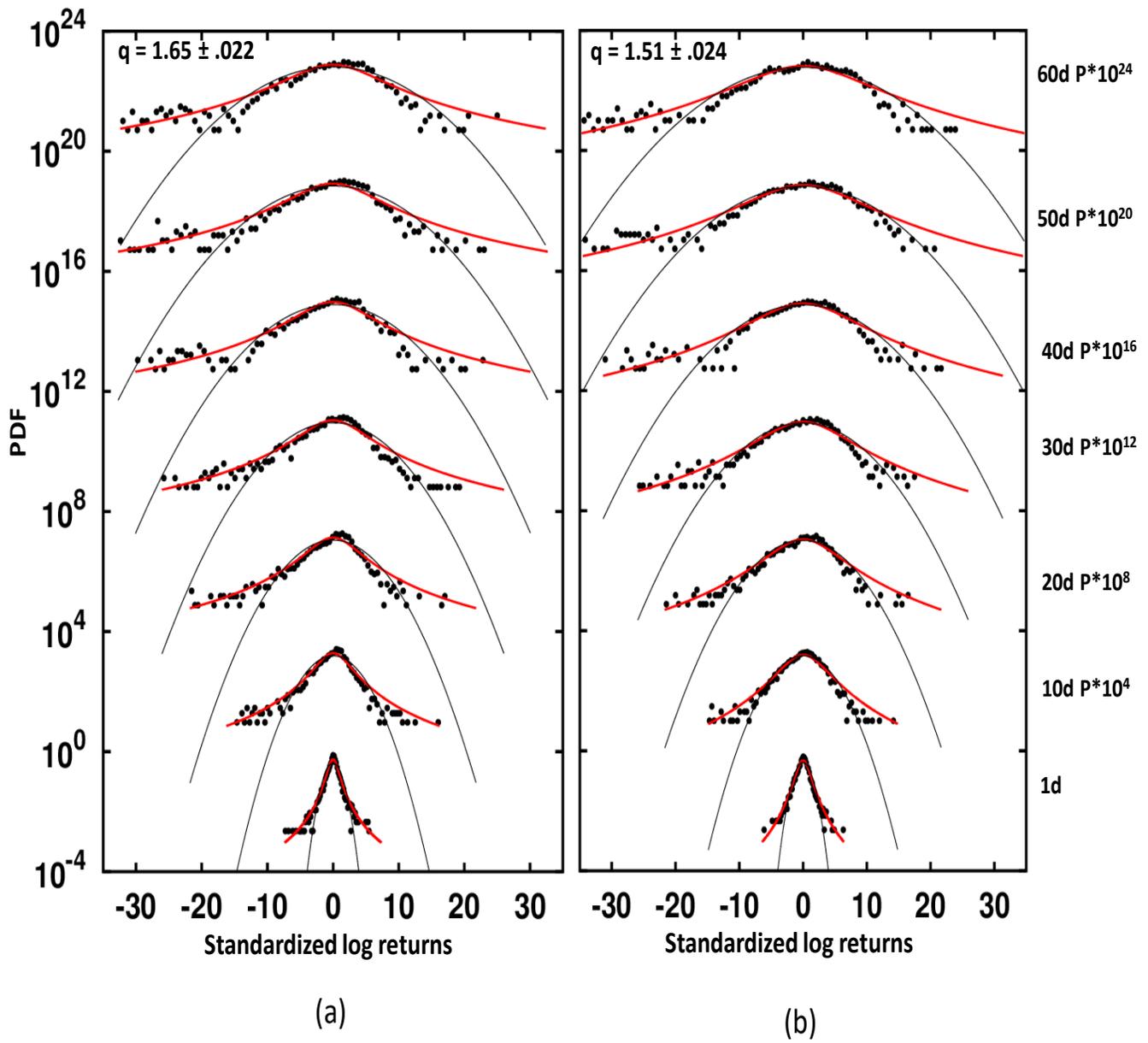

Figure 4.  Same as Figure 3, for region 2 (30 Jul 2002 – 4 Sep 2013).

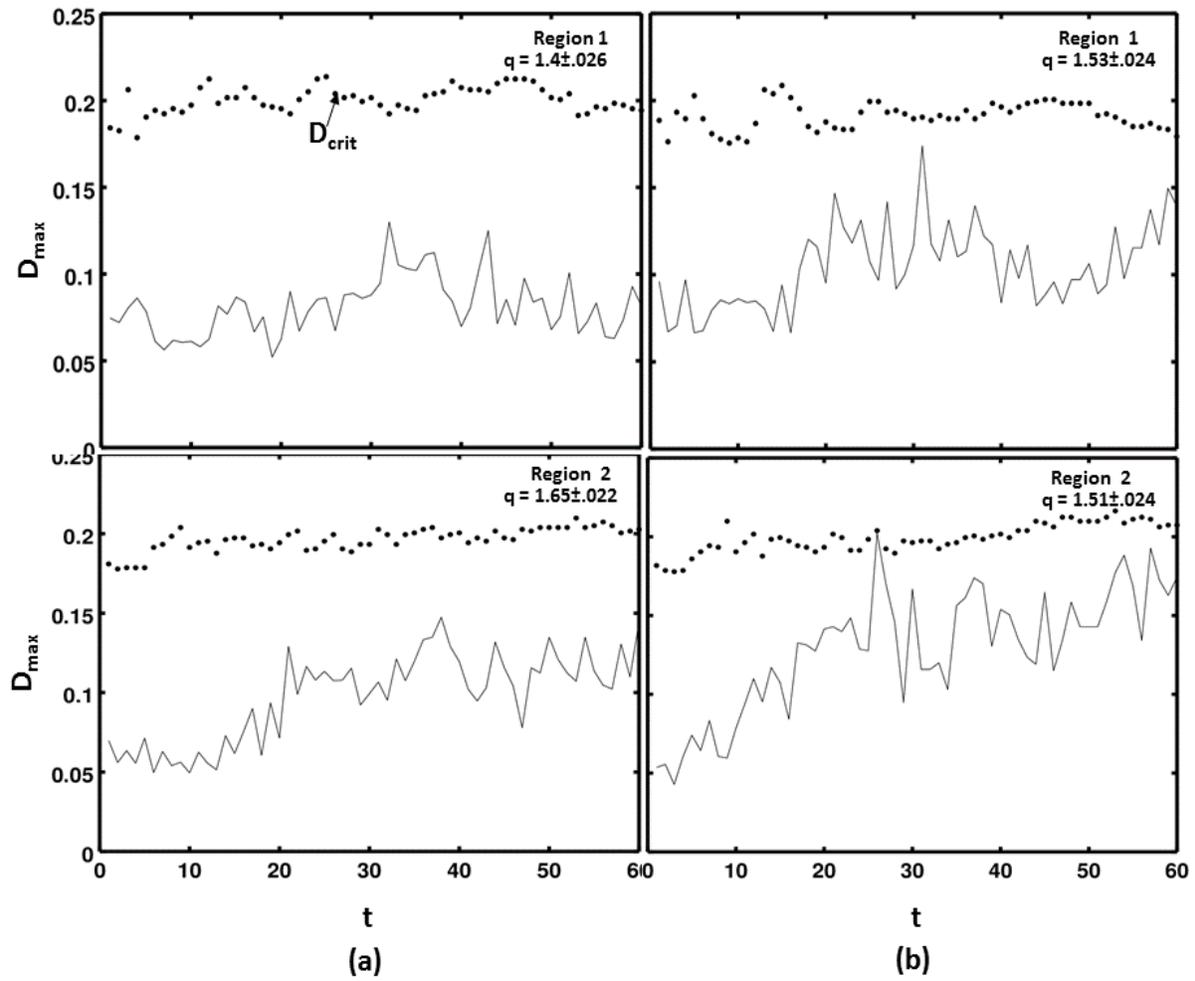

Figure 5. KS goodness of fit test 1. Maximum distance between the empirical and synthetic CDF's are shown as functions of delay in days. (a) S&P 500 and (b) Nasdaq.

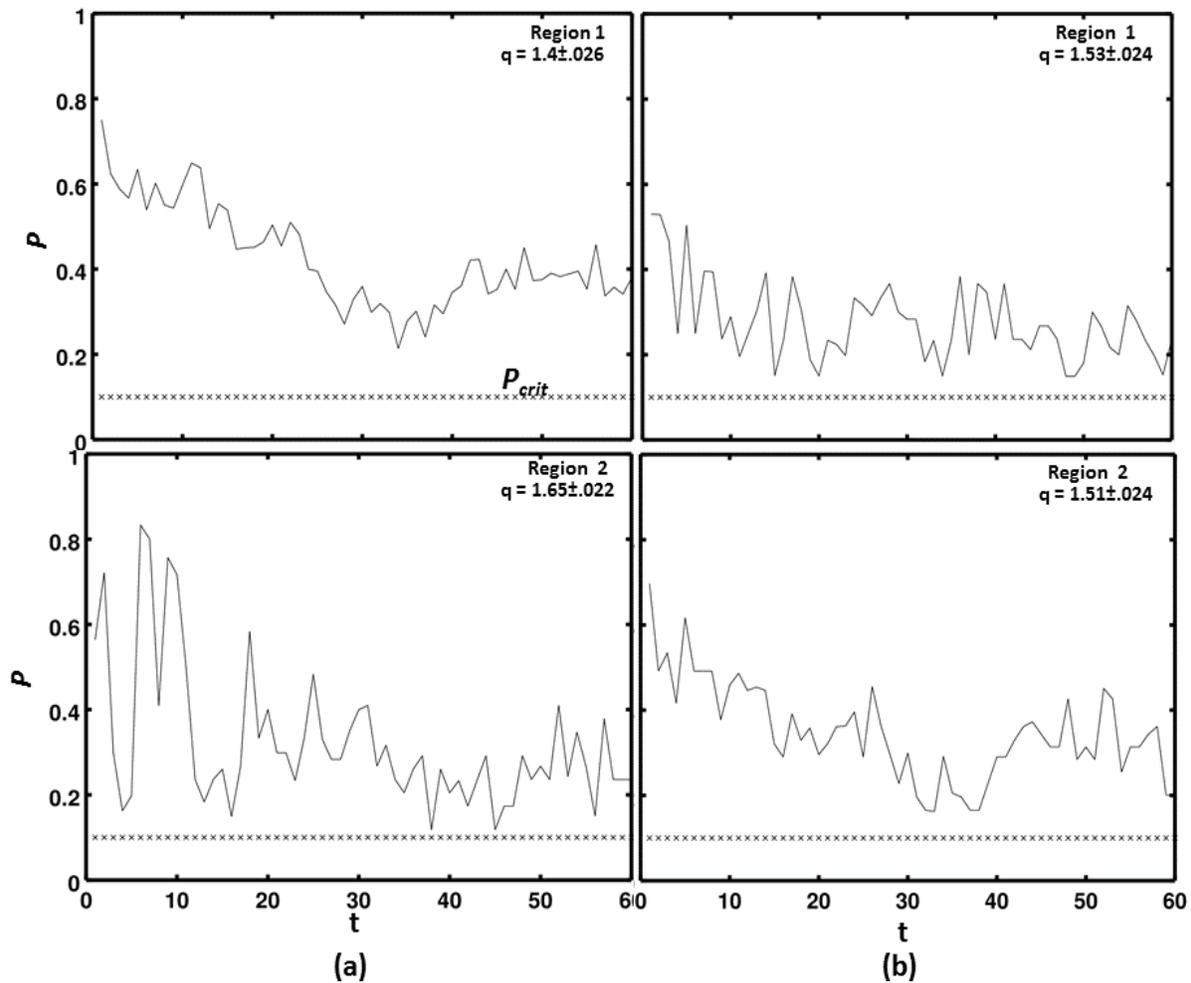

Figure 6. KS goodness of fit test 2. **P** values as functions of delay. (a) S&P 500 and (b) Nasdaq.

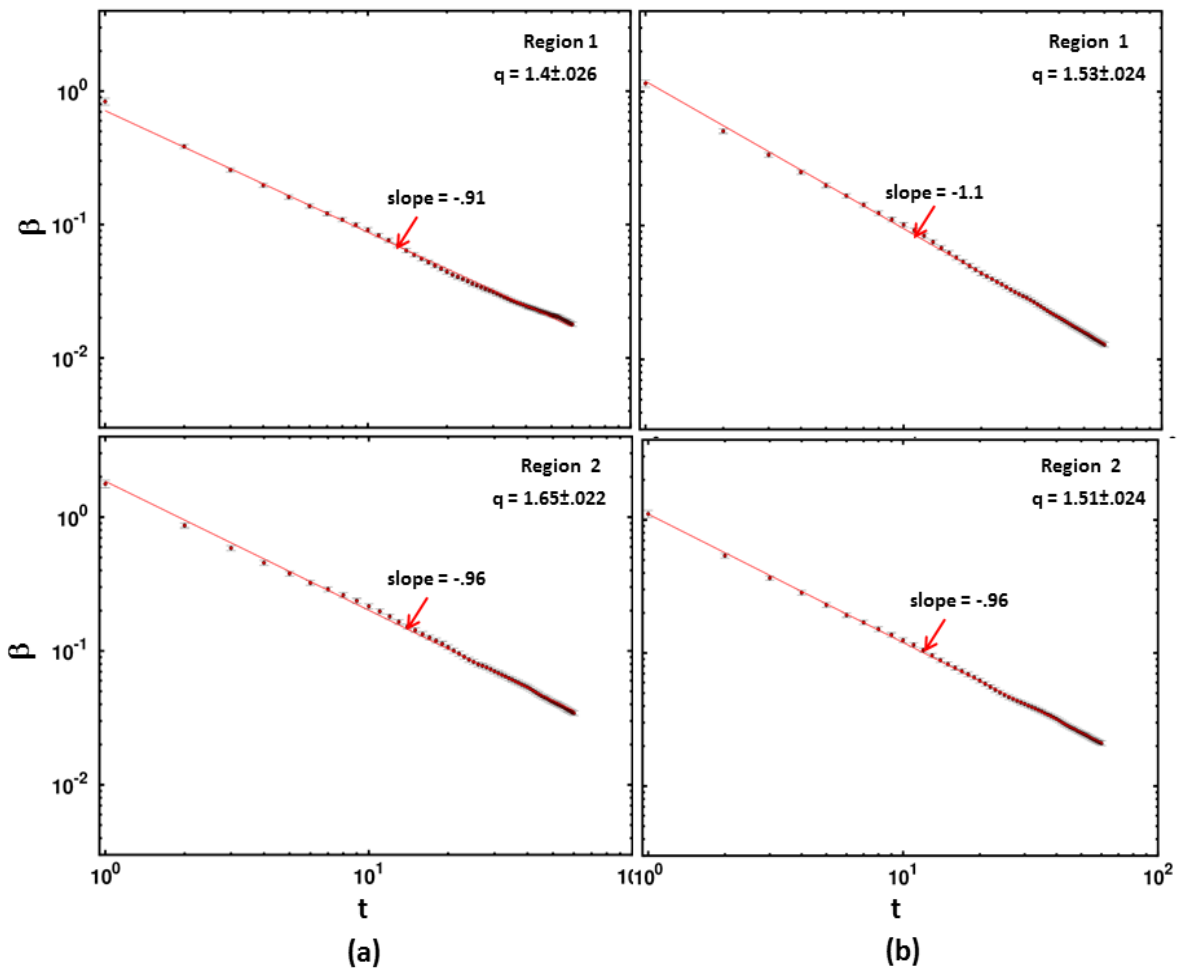

Figure 7. Variation of the estimated $\hat{\beta}$ with the delay $t$ for regions 1 and 2. The error bars for $\hat{\beta}$ are also shown. The solid red line is the linear fit. (a) S&P 500 and (b) Nasdaq.

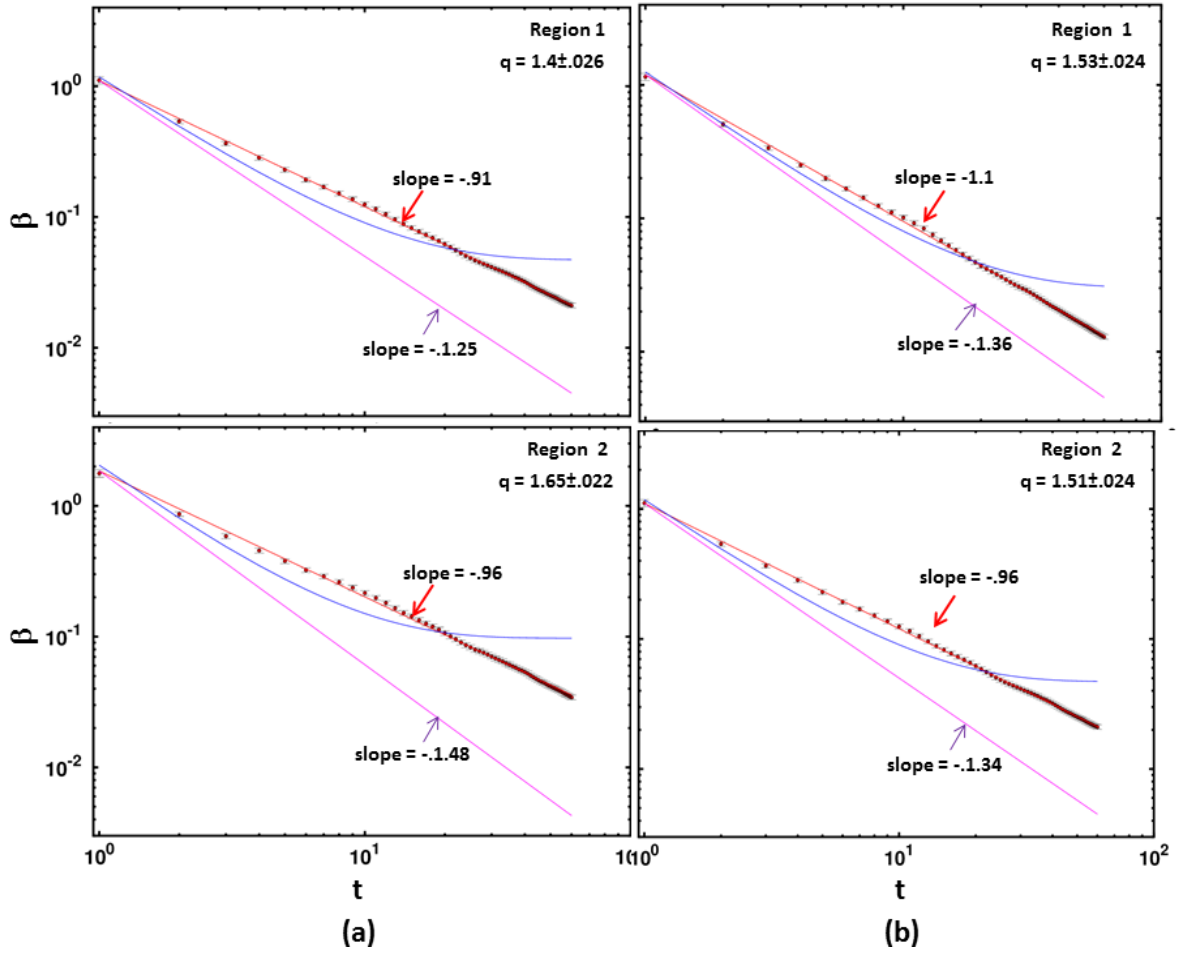

Figure 8. Comparison of $\hat{\beta}$ with $\beta_{dd}$ and $\beta_{sd}$. Red – estimated, Blue – drift + diffusion ($\beta_{dd}$), and Magenta – superdiffusion ($\beta_{sd}$). The solid red line is the linear fit to $\hat{\beta}$ vs. $t$. (a) S&P 500 and (b) Nasdaq.

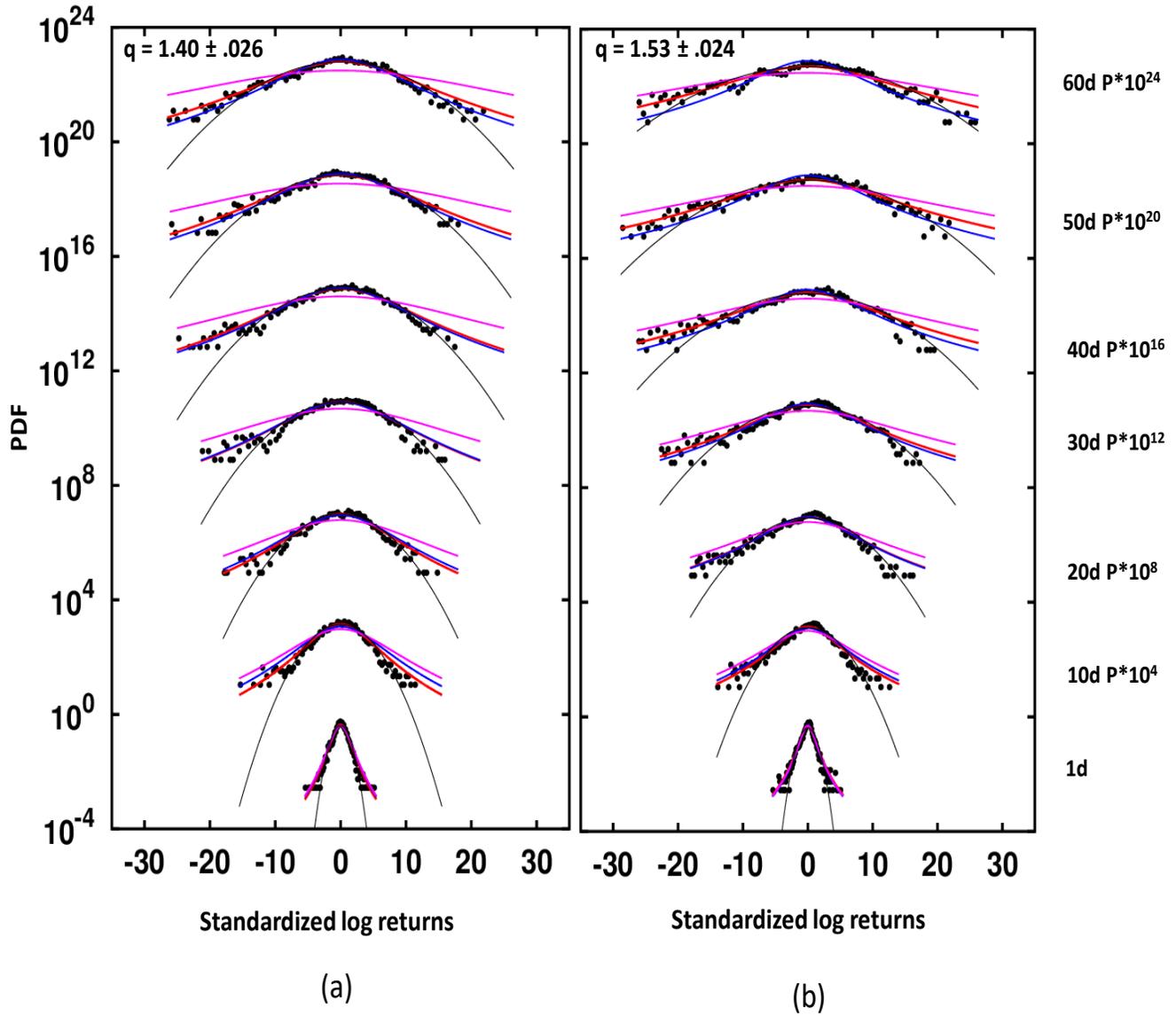

Figure 9. Comparison of the Tsallis distributions from estimated $\hat{\beta}$, $\beta_{dd}$ and $\beta_{sd}$ with the data distributions for region 1. Red – Estimated, Blue – drift + diffusion, Magenta – superdiffusion, and Black – Gaussian. (a) S&P 500 and (b) Nasdaq.

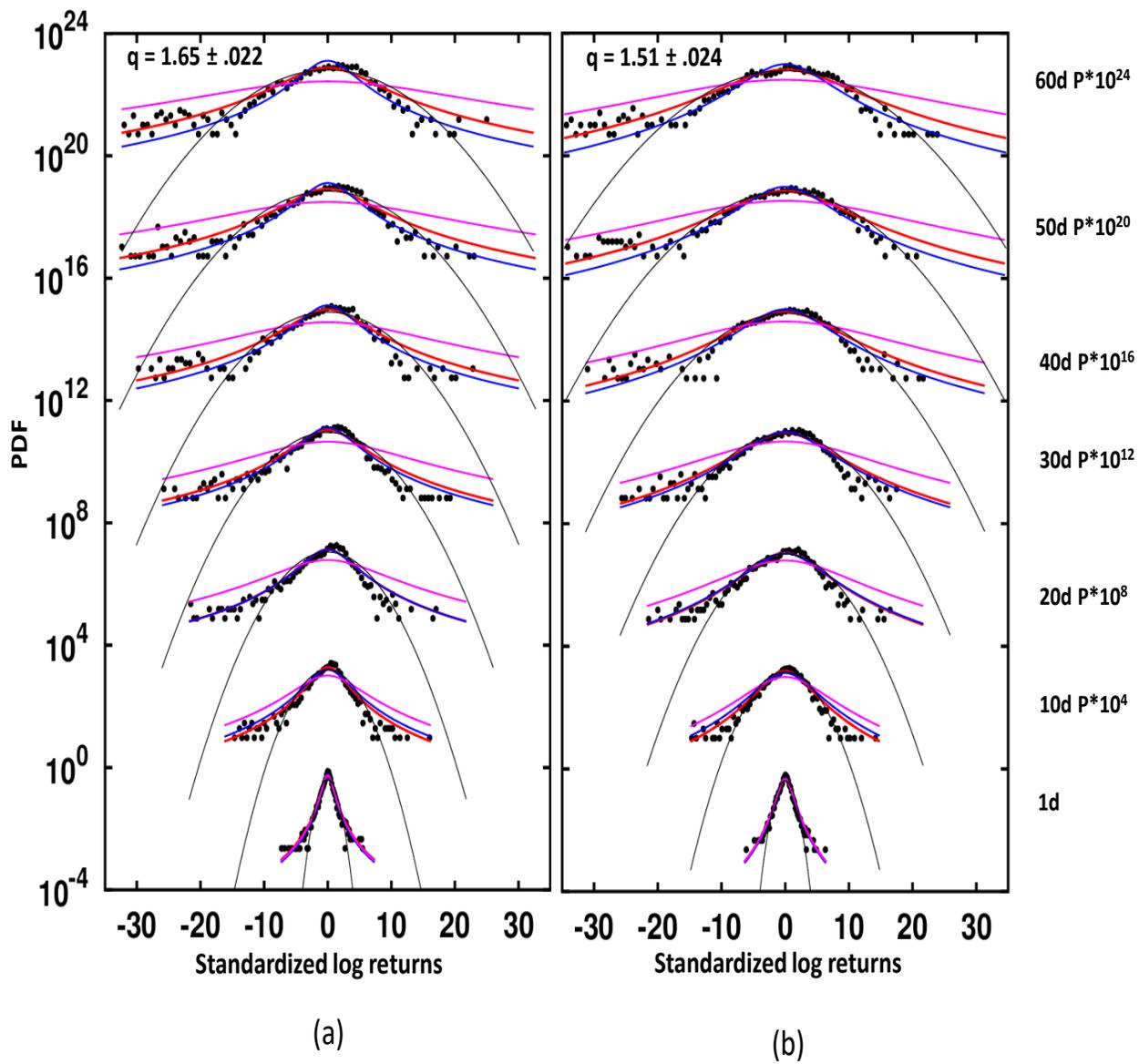

Figure 10. Same as Figure 7 for region 2.

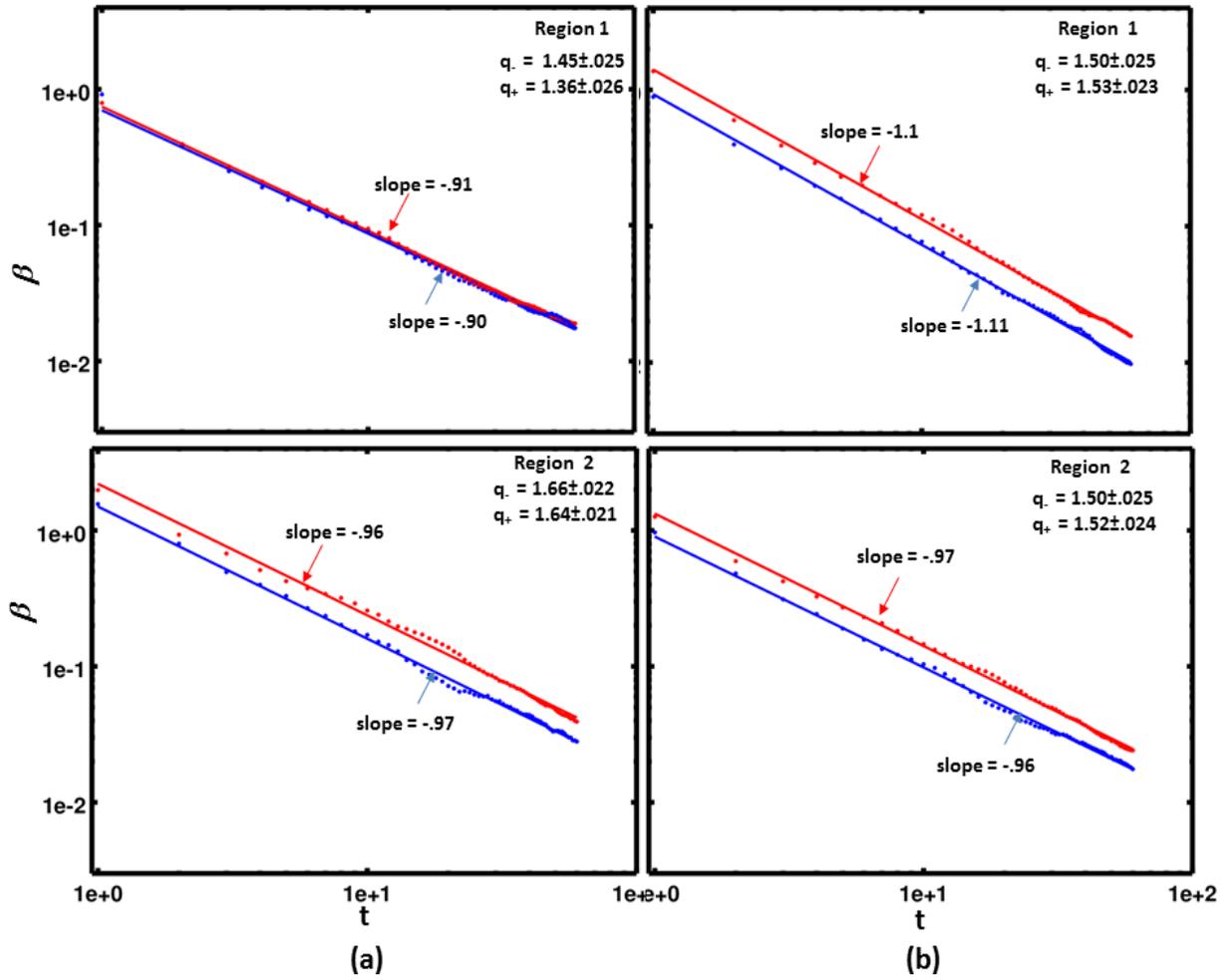

Figure 11. Variation of $\beta$ with delay in the asymmetric case. (a) S&P 500 and (b) Nasdaq. Blue corresponds to the left branch of the distributions (-ve returns) and Red – the right branch (+ve returns).

Table 1.  The estimated values of the drift parameter b, the diffusion parameter D and the characteristic time (in days) $\tau \; = \; 1/(b(3-q))$.

| Index | Region | b | D | Tau |
|---|---|---|---|---|
| S&P 500 | 1 | .043∓.0024 | .506 | 14.31 |
| Nasdaq | 1 | .046∓.0031 | .412 | 14.87 |
| S&P 500 | 2 | .101∓.0068 | .393 | 7.29 |
| Nasdaq | 2 | .064∓.0047 | .423 | 10.47 |